# COMPARISON OF TWO SINGLE-ION OPTICAL FREQUENCY STANDARDS AT THE SUB-HERTZ LEVEL

CHR. TAMM, T. SCHNEIDER, AND E. PEIK

*Physikalisch-Technische Bundesanstalt (PTB),
Bundesallee 100, 38116 Braunschweig, Germany
E-mail: christian.tamm@ptb.de*

We describe experimental investigations on an optical frequency standard based on a laser cooled $^{171}Yb^+$ ion confined in a radiofrequency Paul trap. The electric-quadrupole transition from the $^2S_{1/2}(F=0)$ ground state to the $^2D_{3/2}(F=2)$ state at the wavelength of 436 nm is used as the reference transition. The reference transition is probed by a frequency-doubled, frequency-stabilized diode laser and is resolved with a Fourier-limited full halfwidth of approximately 30 Hz. In order to compare two $^{171}Yb^+$ standards, separate frequency shift and servo systems are employed to stabilise the probe frequency to the reference transition line centers of two independently stored $^{171}Yb^+$ ions. The present experimental results indicate a relative instability (Allan standard deviation) of the optical frequency difference between the two systems of $\sigma_y(1000\ s)=1.0\cdot 10^{-15}$ and a mean frequency difference of 0.2 Hz. Shifts in the range of several Hertz are observed in the frequency difference if a stationary electric field gradient is superimposed on the radiofrequency trap field. This measurement permits a first experimental estimate of the electric quadrupole moment of the $^2D_{3/2}$ state of $Yb^+$.

## 1 Introduction

$^{171}Yb^+$ is an attractive candidate for optical frequency standards based on a trapped, laser-cooled single ion because reference transitions with vanishing low-field linear Zeeman frequency shift are available in a level system with relatively simple hyperfine and magnetic sublevel structure [1,2]. The electric-quadrupole transition $^2S_{1/2}(F=0)$ - $^2D_{3/2}(F=2)$ of $^{171}Yb^+$ is at a wavelength of 436 nm and has a natural linewidth of 3.1 Hz. The absolute optical frequency of this transition was measured with a total $1\sigma$ fractional uncertainty of $1\cdot 10^{-14}$, so that it is now one of the most accurately known atomic transition frequencies in the optical wavelength range [3,4].

Here we present experimental results on the high-resolution spectroscopy of the 436 nm reference transition of $^{171}Yb^+$, and initial results on the comparison of two $^{171}Yb^+$ optical frequency standards. With respect to the statistical uncertainty of the comparison and the ability to resolve small frequency offsets, the results yield an improvement by more than one order of magnitude over previous related work [5]. Quadrupole shifts of the atomic transition frequency of the order of a few Hertz which are introduced in one trap by superimposing a constant component on the confining radiofrequency field can be clearly resolved. This permits a first experimental estimate of the electric quadrupole moment of the $^2D_{3/2}$ state of $Yb^+$. The quadrupole shift caused by electric stray fields is expected to be one of the largest systematic frequency shift effects in optical frequency standards that use ions with alkali-like level systems such as $^{171}$Yb$^+$ and $^{199}$Hg$^+$ [6].

## 2  Optical-excitation scheme

A scheme of the lowest-lying energy levels of $^{171}$Yb$^+$ is shown in Fig. 1. For laser cooling, the low-frequency wing of the quasi-cyclic F=1 - F=0 component of the $^2$S$_{1/2}$ - $^2$P$_{1/2}$ resonance transition is excited, and a static magnetic field of approximately 300 µT is applied in order to prevent optical pumping to a nonabsorbing superposition of the magnetic sublevels of the F=1 ground state. The natural linewidth of the resonance transition is 21 MHz, which implies a one-dimensional kinetic temperature of 0.6 mK at the Doppler cooling limit. A weak sideband of the cooling radiation provides hyperfine repumping from the F=0 ground state to the $^2$P$_{1/2}$(F=1) level. At the end of each cooling phase, the hyperfine repumping is switched off in order to prepare the ion in the F=0 ground state.

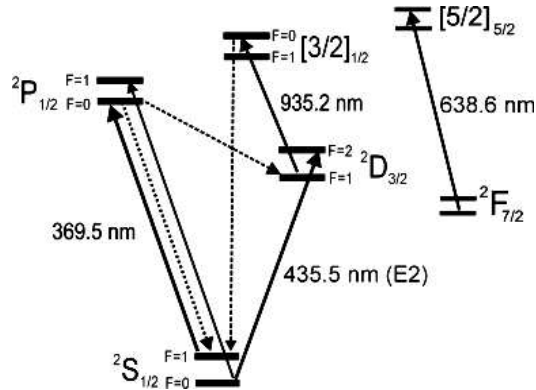

Fig. 1: Low-lying energy levels of $^{171}$Yb$^+$ and optical excitation scheme. The main spontaneous decay paths are indicated by dashed lines. Hyperfine splittings are not drawn to scale. The hyperfine splitting frequencies of the S, P, D, and [3/2] levels are respectively given by 12.6 GHz, 2.1 GHz, 0.9 GHz, and 2.5 GHz.

The rapid spontaneous decay from the $^2$P$_{1/2}$ state to the metastable $^2$D$_{3/2}$(F=1) level that occurs during laser cooling is compensated for by coupling this level to the [3/2]$_{1/2}$(F=1) state, from where the ion readily returns to the ground state. The extremely long-lived $^2$F$_{7/2}$ state, which is populated at a rate of ≈ 0.3 h$^{-1}$, is depleted by excitation to the [5/2]$_{5/2}$ level.

The F=2 sublevel of the $^2D_{3/2}$ state is not rapidly populated or depleted by the laser cooling excitation. Individual quantum jumps to this state due to excitation of the reference transition can therefore be detected through the interruption of the resonance fluorescence scattering.

In the experiments described below, the cooling and reference transitions are excited alternately in measurement cycles of 90 ms duration. During the excitation of the reference transition, the magnetic field is decreased to the microtesla range in order to reduce the quadratic Zeeman frequency shift. When observing the absorption spectrum of the reference transition, the excitation probability to the $^2D_{3/2}(F=2)$ state is registered as a function of the probe laser detuning. In order to operate the system as a frequency standard, both wings of the central resonance of the absorption spectrum are probed alternately, and the probe light frequency is stabilised to the line center according to the difference of the measured excitation probabilities.

## 3  Experimental setup

The employed ion traps are cylindrically symmetric with a ring electrode diameter of 1.4 mm. Except for the measurement of the quadrupole shift of the $^2D_{3/2}$ level, the applied trap drive voltage contained no constant component. In this case, the axial and radial secular motion frequencies of a trapped Yb$^+$ ion are in the range of 0.7 MHz and 1.4 MHz, respectively. Typical single-ion storage times are in the range of several months. Static electric stray fields in the confinement volume are compensated in three dimensions so that the amplitude of stray-field induced micromotion is smaller than the secular motion amplitude at the Doppler cooling limit.

Cooling radiation at 370 nm is generated by frequency doubling the output of an extended-cavity diode laser. Hyperfine repumping radiation is produced by modulating the injection current of this diode laser at a frequency near 14.7 GHz. Extended-cavity diode lasers are also used to generate 935 nm and 639 nm repumping radiation, and light at 871 nm which is frequency doubled in order to produce the 436 nm probe radiation. The cooling and repumping radiation is blocked by mechanical shutters during the excitation of the reference transition.

In order to stabilise the frequency of the 436 nm probe light, a Pound-Drever-Hall scheme is used to lock the 871 nm diode laser with a servo bandwidth of 0.5 MHz to a fiber-coupled high-finesse ULE cavity. The cavity is suspended in vacuum by springs of 1 m length for vibration isolation. The cavity temperature is actively stabilised so that the drift of the 436 nm probe frequency is typically mainly determined by the 0.07 Hz/s long-term aging drift of the cavity material [7].

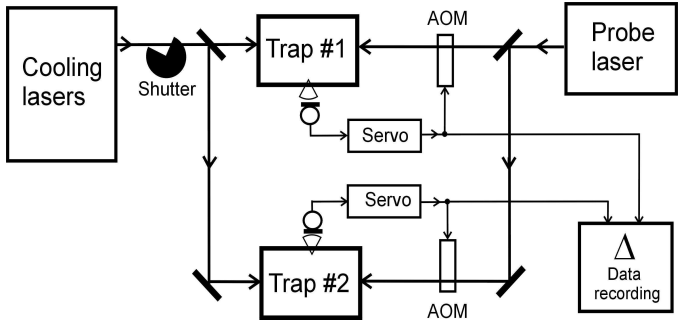

Fig. 2. Experimental setup for comparison of two $^{171}$Yb$^+$ frequency standards. AOM: acoustooptic modulators, providing independent frequency shifts between the probe laser and the two ion traps.

The scheme of the frequency comparison experiment is shown in Fig. 2. Both traps use the same cooling laser setup and synchronous timing schemes for cooling, state preparation, and state detection. Using two independent digital servo systems, the error signals resulting from the probing of the atomic resonances are averaged over typically 20 measurement cycles before the detunings between the probe laser frequency and the probe light beams incident on the traps are corrected. In order to minimise servo errors due to the drift of the probe laser frequency, a second-order integrating servo algorithm is used. The servo time constants are in the range of 30 s. The differences of the detunings imparted on the probe beams are averaged over time intervals of 1 s and recorded.

## 4  Spectroscopy of the reference transition

Absorption spectra of the $^2S_{1/2}(F=0) - ^2D_{3/2}(F=2)$ transition of a single trapped $^{171}$Yb$^+$ ion are shown in Fig. 3. They were obtained using the setup described in Ref. 7. In Fig. 3, the frequency resolution increases from Fig. 3(a) to (d). For Fig. 3(a) and (b), the linewidth of the probe radiation was increased by white-noise frequency modulation in order to reduce the number of data points required for the scan. Due to the incoherent optical excitation, the absorption probability here is limited to 0.5 at full saturation. Figure 3(a) shows that the strength of the radial secular motion sidebands is significantly smaller than that of the central recoilless component. This permits the conclusion that the Lamb-Dicke condition is well satisfied for the excitation of the reference transition. In Fig. 3(b), the Zeeman structure of the recoilless component is shown for an applied static magnetic field of approximately 1.1 µT. Magnetic fields in the range of 1 µT were also applied in the frequency comparison experiments described below. Figures 3(c) and (d) show the central $\Delta m_F=0$-component of (b) for the case of coherent π-pulse excitation and approximately Fourier-limited resolution. In Fig. 3(d), the maximum absorption probability is reduced relative to Fig. 3(c) because the short-time fluctuations of the probe laser frequency are not negligible relative to the Fourier linewidth limit of 27 Hz.

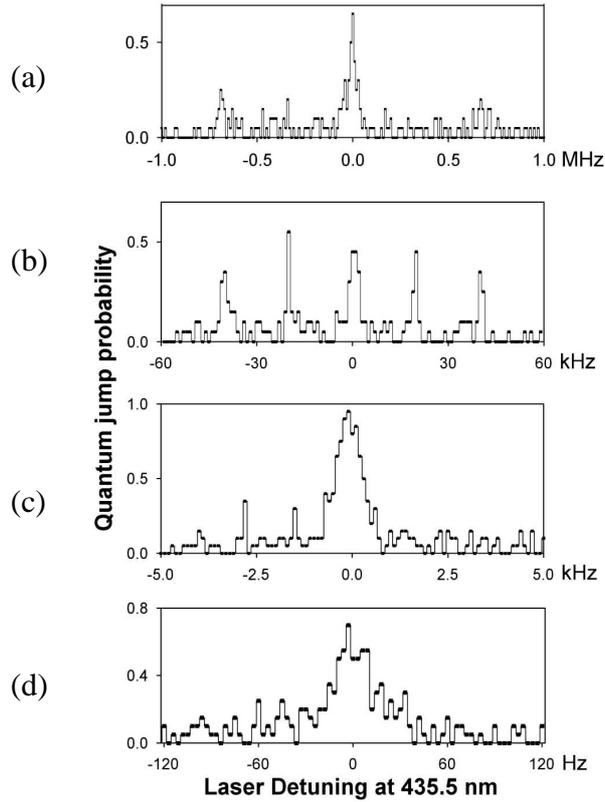

Fig. 3: Absorption spectra of the $^2S_{1/2}(F=0)$ - $^2D_{3/2}(F=2)$ transition of a single trapped $^{171}Yb^+$ ion, showing the first-order radial secular-motion sidebands and the central carrier resonance (a); the Zeeman pattern of the carrier resonance (b); and the $\Delta m_F=0$-resonance of (b) in higher resolution (c, d). Each data point corresponds to an average of 20 measurement cycles. The probe pulse length was 1 ms in (a), (b), and (c), and 30 ms in (d). For further details see text.

## 5   Absolute transition frequency and systematic frequency shifts

Using a femtosecond frequency comb generator, the frequency of the 436 nm $^2S_{1/2}$(F=0) - $^2D_{3/2}$(F=2, $m_F$=0) transition of $^{171}$Yb$^+$ was measured relative to a caesium fountain microwave frequency standard [3]. The optical-excitation conditions were identical to those of Fig. 3(d).

The measured absolute frequency is $\nu_{Yb+}$= 688 358 979 309 312 ± 6 Hz. This frequency value includes the shift of the transition frequency due to isotropic blackbody radiation at an ambient temperature of 300 K. On the basis of computed atomic oscillator strengths, the shift is calculated as -0.4 Hz [8]. The total 1 $\sigma$ measurement uncertainty of ± 6 Hz consists of approximately equal statistical and systematic contributions. The dominant source of the systematic measurement uncertainty is given by the electric-quadrupole interaction of the upper level of the reference transition with the gradient of stationary electric stray fields. A maximum stray-field induced quadrupole shift of the order of 1 Hz is expected for atomic $D_{3/2}$ and $D_{5/2}$ states [9]. A non-negligible systematic uncertainty contribution also arises from servo errors due to drifts of the probe laser frequency. The uncertainty contributions of other frequency shifting effects are negligible under the present experimental conditions. The magnetic field applied during the excitation of the reference transition leads to a quadratic Zeeeman shift of only 0.05 Hz. Since the trapped ion is cooled to the Doppler limit, the second-order Doppler and Stark effect shifts caused by the trap field are expected to be in the millihertz range [6].

## 6   Comparison between two traps

Figure 4 shows the temporal variation of the frequency difference between two independent $^{171}$Yb$^+$ trap and servo systems, using the experimental setup shown in Fig. 2. The Allan deviation of this data set is shown in Fig. 5. The conditions of this measurement were similar to those of Fig. 3(d).

Using temporally overlapping probe pulses, the atomic resonance signals were resolved with nearly Fourier-limited linewidths of approximately 30 Hz in both traps. The mean frequency difference of the data shown in Fig. 4 is $\langle\Delta\rangle \approx$ 0.2 Hz, corresponding to a relative optical frequency offset of $3\cdot10^{-16}$. Since $\langle\Delta\rangle$ is smaller than the Allan deviation for long averaging times ($\sigma_y(\tau) \approx 1\cdot10^{-15}$ for $\tau \geq$ 800 s), the observed offset is not statistically significant. A change of the drive voltage amplitude of one of the traps by 15% did not cause any significant frequency offset at the 1 Hz level.

As shown in Fig. 5, the variation of the Allan deviation with the averaging time $\tau$ is in qualitative agreement with a numerical calculation which simulates the effect of quantum projection noise for the realized experimental conditions. The observed Allan deviation however exceeds the quantum projection noise limit by approximately a factor of two. A possible reason for this excess instability are temporal fluctuations of the probe laser emission spectrum which can lead to fluctuating servo errors. The frequency shifts caused by this effect are not necessarily equal for both servo systems because the probe pulse areas by which the two ions were excited were not exactly matched.

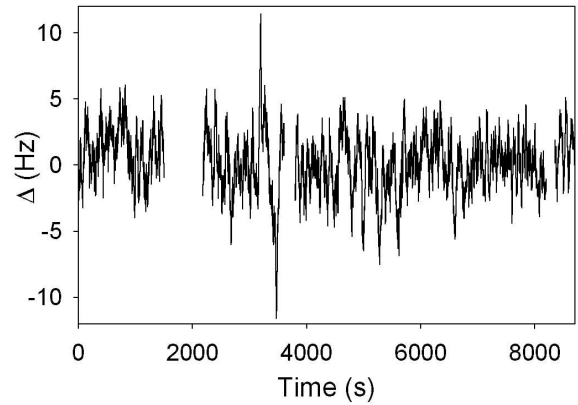

Fig. 4: Temporal variation of the frequency difference between two probe light fields independently frequency stabilised on the 436 nm reference transitions of two trapped $^{171}$Yb$^+$ ions. The average frequency difference calculated from this data set is 0.2 Hz. The intervals without data points correspond to times when no frequency correction signal was produced by one of the trapped ions.

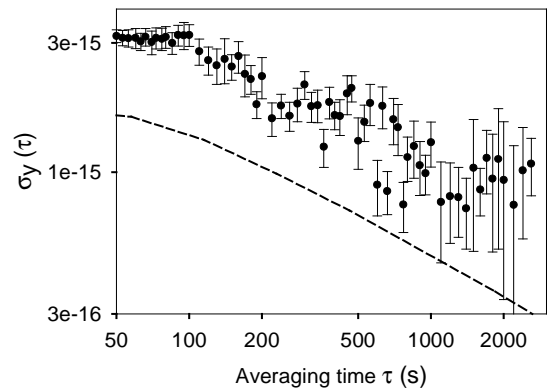

Fig. 5: Allan standard deviation of the data set shown in Fig. 4, normalized to the optical frequency of 688 THz. The dashed line shows the result of a Monte Carlo simulation of the servo action for the case that the fluctuations of the atomic resonance signals are determined by quantum projection noise.

## 7 Quadrupole shift measurement

One expects that the interaction of the quadrupole moment of the $^2D_{3/2}$ state of $^{171}Yb^+$ with a static electric field gradient leads to a shift of the frequency of the 436 nm reference transition. In order to experimentally determine the $^{171}Yb^+$ quadrupole moment, a static field gradient was generated in one of the traps by superimposing a constant (dc) voltage on the radiofrequency trap drive voltage. The orientation of this field gradient is determined by the symmetry axis of the trap. The other trap was operated with a pure rf voltage and served as a reference. The result of a corresponding frequency comparison measurement is shown in Fig. 6.

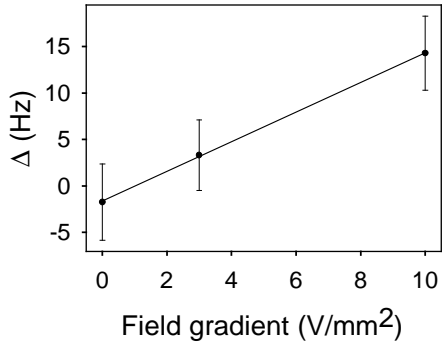

Fig. 6: Frequency difference between the 436 nm light fields stabilised to two $^{171}Yb^+$ traps as a function of the dc field gradient generated in one of the traps. The application of a positive voltage to the endcap electrodes increases the optical frequency. The error bars show the statistical measurement uncertainty.

A three-axis magnetic field sensor was used to determine the orientation of the static magnetic field relative to the applied field gradient. Using the formalism described in Ref. 9, we infer a quadrupole moment of $\Theta = (3.9 \pm 1.9)ea_0^2$ for the $^2D_{3/2}$ level of $Yb^+$ with $e$ being the electron charge and $a_0$ the Bohr radius. The uncertainty of the inferred $\Theta$ value is mainly determined by the uncertainty of the measurement of the angle between the magnetic field and the trap axis.

**References**

[1] Chr. Tamm, D. Engelke, and V. Bühner, *Phys. Rev. A* **61**, 053405 (2000).
[2] S.A. Webster, P. Taylor, M. Roberts, G.P. Barwood, and P. Gill,
    *Phys. Rev. A* **65**, 052501 (2002).

[3] J. Stenger, Chr. Tamm, N. Haverkamp, S. Weyers, and H.R. Telle,
*Opt. Lett.* **26**, (2001).
[4] T. Quinn, *Metrologia* **40**, 103 (2003).
[5] G. Barwood, K. Gao, P. Gill, G. Huang, and H.A. Klein,
*IEEE Trans.* IM-**50**, 543 (2001).
[6] see, e.g., A. Bauch and H.R. Telle, *Rep. Prog. Phys.* **65**, 789 (2002).
[7] Chr. Tamm, T. Schneider, and E. Peik, in: *Proceedings of the 6th Symposium on Frequency Standards and Metrology*, ed. P. Gill (World Scientific, Singapore 2002), p. 369.
[8] B. C. Fawcett and M. Wilson, *At. Data Nucl. Data Tabl.* **47**, 241 (1991); J.W. Farley and W.H. Wing, *Phys. Rev. A* **23**, 2397 (1981).
[9] W.M. Itano, *J. Res. NIST* **105**, 829 (2000).